\documentclass[twocolumn,showpacs,prl]{revtex4}
\usepackage{graphicx}

\usepackage{bm}
\usepackage{amsmath}
\usepackage{amsfonts}
\newcommand{\ch}{{\cal H}}

\newcommand{\tr}{{\rm Tr}}
\newcommand{\ket}[1]{| #1 \rangle}
\newcommand{\bra}[1]{\langle #1 |}

\begin{document}
\title{
Security of  
quantum key distribution with discrete
rotational symmetry}

\author{Masato Koashi}
\affiliation{Division of Materials Physics, Graduate School
of Engineering Science, Osaka University, 1-3 Machikaneyama, 
Toyonaka, Osaka 560-8531, Japan}
\affiliation{CREST Photonic Quantum Information Project, 4-1-8 Honmachi, Kawaguchi, Saitama 331-0012, Japan}

\begin{abstract} 
We prove the unconditional security of quantum key distribution protocols 
using attenuated laser pulses with $M$ different linear polarizations.
When $M=4$, the proof covers the so-called SARG04 protocol 
[V.~Scarani et al., Phys.\  Rev.\ Lett.\ {\bf 92}, 057901 (2004)],
which uses exactly the same quantum communication as the 
Bennett-Brassard 1984 protocol. For a channel with transmission 
$\eta$, we show that the key rate in SARG04 scales as $O(\eta^{3/2})$.
When we increase the number of states to $M=2k-1$ or $2k$, 
the key rate scaling improves as $O(\eta^{(k+1)/k})$.

\end{abstract}

\maketitle

Information encoded on the polarization of
a single photon is strongly affected by 
the law of quantum mechanics, and can be 
used to grow a shared random 
bit sequence (secret key) 
between two remote parties 
with negligible leak to an eavesdropper.
The first protocol of this 
quantum key distribution (QKD) was
proposed by Bennett and Brassard \cite{Bennett-Brassard84}, 
and is called the BB84 protocol. A practically 
important and rather surprising fact is 
that the QKD is still possible \cite{ILM01} 
even if we encode the information on an attenuated 
laser pulse, which might be regarded as a 
classical object. One drawback in using 
such a practical light source
 is the decrease of the efficiency when we 
take the channel loss into consideration \cite{Lutkenhaus00,BLMS00}. 
For a channel with transmission $\eta$,
the secure key rate in the 
BB84 protocol scales as $O(\eta^2)$ 
instead of naively expected $O(\eta)$ dependence.
The reason of this poor performance is that
whenever the sender Alice emits more than 
one photon, the eavesdropper Eve can keep 
extra photons without introducing any 
detectable error. In order to suppress 
this so-called photon number splitting (PNS)
attack, 
Alice must attenuate her laser pulse 
in proportion to $\eta$.

A number of possibilities have been studied 
to remedy this performance drop. 
One solution is to switch to a completely different
protocol and implement the B92 protocol 
with a strong phase-reference pulse 
as in its original proposal \cite{Bennett92}. 
It was proved that 
this simple protocol using laser light
achieves the key rate of $O(\eta)$ \cite{Koashi04}.
Another solution is to modify the quantum 
communication part of the BB84 protocol to 
detect the PNS attack. 
This can be done \cite{Hwang03} by mixing the pulses 
(decoy states) with 
various amplitudes in the protocol, and 
it was proved \cite{LMC05} that the key rate of $O(\eta)$
can be achieved with a large number of decoy states.
A third possibility is the so-called SARG04 protocol \cite{SARG04}, 
which modifies {\em only} the classical communication 
part of the BB84. Since the feasibility of the BB84
with attenuated laser pulses has been 
repeatedly tested experimentally in the past decade,
the SARG04 protocol has its unique practical importance.
What we know about its security so far is the following.
Tamaki and Lo have analyzed \cite{Tamaki-Lo04}
a protocol which is 
the SARG04 augmented with decoy states, and have shown 
that unconditionally secure key can be extracted from the 
two-photon emission part. For the unmodified SARG04 with 
its original spirit, Branciard {\em et al}. derived 
$O(\eta^{3/2})$ dependence of the key rate assuming a 
limited set of individual attacks by Eve \cite{BGKS05}.

In this paper, we prove the security of the SARG04 
protocol and its natural generalization to $M$-state
protocols with no condition on the attack by 
an eavesdropper. We show a lower bound on the key rate in 
the SARG04 protocol scaling as $\sim \alpha \eta^{3/2}$,
where $\alpha$ is a factor depending on the bit error rate.
For $M(> 4)$ states, the exponent improves to 
$O(\eta^{(k+1)/k})$ with $k=\lceil M/2 \rceil$, while 
 the requirement for the bit error rate becomes severer
as $k$ increases. For the security proof, we use the 
$2M$-fold discrete rotational symmetry of the system to 
simplify the argument. For a light source with any 
photon-number distribution, the protocol can be reduced to 
an entanglement based protocol with Hilbert space
$\ch\cong\mathbb{C}^M\otimes \mathbb{C}^2$.
The whole space $\ch$ can be further divided into $M$ qubits 
according to the angular momentum.
Then we can use a 
standard analysis to obtain the security proof.

We consider the following protocols specified by 
two integers $(M,L)$ with $2L\le M$.
Alice prepares system $C$ in a
 linearly polarized, phase-randomized state of light, which is 
written as
$$
\hat\rho(\theta)\equiv \sum_n \mu_n \ket{\theta,n}_C{}_C\bra{\theta,n},
$$
$$
\ket{\theta,n}_C\equiv  2^{-n/2} (n!)^{-1/2} 
(e^{i\theta}\hat{a}^\dagger_{-1}+e^{-i\theta}\hat{a}^\dagger_{1})^n\ket{vac}_C,
$$
where $\hat{a}_k$ is the annihilation operator for a circularly 
polarized photon with angular momentum $k$, and $\mu_n$ stands for 
the photon number distribution. The angle $\theta$ is chosen 
from the set $\Omega_M\equiv \{\pi l/M: l=0,\ldots, M-1\}$.
Alice encodes her randomly chosen 
bit $a$ in the following way.
She chooses an integer $j (0\le j \le M-1)$ randomly,
and sends $\hat\rho(a\Theta+\pi j/M)$ to Bob, where
$$
\Theta\equiv \pi L/M.
$$

The receiver Bob analyzes the polarization of the 
received light by a polarization beam splitter (PBS)
followed by two photon detectors $D_1$ and $D_2$.
The whole analyzer is rotated by a randomly chosen angle
$\theta' \in \Omega_M$, such that the light in 
state $\hat\rho(\theta')$ would be directed to $D_1$ and 
never to $D_2$.
We assume that 
the dark counts and the inefficiency of the detectors 
can be equivalently described by a noise source in 
the channel, and hence we treat them as ideal detectors.
We say an event is `detected' when $D_1$ and $D_2$
register exactly one photon in total,
implying that the incoming state is found in 
the two-dimensional Hilbert space $\ch_B$ of a single photon.
Bob announces when the event is detected, and 
records the rate $\eta_d$ of detected events, 
which we call the detection  rate.
After Bob receives the light, Alice announces the value of $j$.
Bob's measurement determines a conclusive value for his bit $b$
only when (a) $D_1$ registers no photons 
and $D_2$ registers one photon, and (b) the analyzer angle satisfies
$\theta'=\Theta+\pi j/M$ ($b=0$ in this case) or $\theta'=\pi j/M$
($b=1$). Let $\ket{k}_B\in \ch_B (k=1,-1)$ be the state of 
a single photon 
with angular momentum $k$, and define 
$\ket{\xi_{\bar{\theta}}}_B\equiv
(e^{i\theta}\ket{-1}_B- e^{-i\theta}\ket{1}_B)/\sqrt{2}$.
Conditioned on the value of $j$, the POVM elements $\hat{B}^{(j)}_b$
for conclusive events ($b=0,1$) are given by 
$\hat{B}^{(j)}_0=P(\ket{\xi_{\overline{\Theta+\pi j/M}}}_B)/M$ 
and 
$\hat{B}^{(j)}_1=P(\ket{\xi_{\overline{\pi j/M}}}_B)/M$,
where $P(\ket{\cdot})\equiv \ket{\cdot}\bra{\cdot}$.
Finally, Bob announces whether the event has been conclusive or not.

It will be obvious that the $(M,L)=(4,2)$ protocol 
is essentially the BB84 protocol, in which the bit $a$
is encoded to orthogonal states ($\Theta=\pi/2$).
The SARG04 corresponds to $(4,1)$ and the bit is 
encoded to nonorthogonal states ($\Theta=\pi/4$), 
which is the crux of the protocol.

The security proof begins with the introduction of
a virtual system $A$ with $M$-dimensional Hilbert space
$\ch_A$. We take a basis $\{\ket{2k}_A\}_{k=0,\ldots,M-1}$
and assume that the state $\ket{2k}_A$ has angular momentum 
$2k$. Let us define 
\begin{multline}
 \ket{\Phi_n}_{AC}\equiv 
\sqrt{2^{-n}n!} \sum_{k=0}^{n} [k!(n-k)!]^{-1}
\\
\ket{2(k\;\text{mod}\;M)}_A\otimes (\hat{a}^\dagger_1)^{n-k}
(\hat{a}^\dagger_{-1})^k
\ket{vac}_C
\nonumber
\end{multline}
and let $\hat{\rho}_{AC}\equiv \sum_n \mu_n P(\ket{\Psi_n}_{AC})$.
This state is invariant under the rotation of systems $AC$
by discrete angles $\theta\in \Omega_M$.
Let us define another orthonormal basis 
$\{\ket{\xi_\theta}_A\}(\theta\in \Omega_M)$ by 
$$
\ket{\xi_\theta}_A\equiv M^{-1/2}
\sum_{k=0}^{M-1} e^{-2i k \theta}\ket{2k}_A.
$$
It is straightforward to see that 
${}_A\bra{\xi_\theta}\hat\rho_{AC}\ket{\xi_\theta}_A=\hat\rho(\theta)$
for $\theta\in \Omega_M$. Therefore, 
it makes no difference if Alice prepares $\hat\rho_{AC}$, 
sends system $C$ to Bob, and then determines $(a,j)$ by 
a measurement with POVM 
$\hat{A}_{a,j}\equiv P(\ket{\xi_\theta}_A)/2$ 
with $\theta=a \Theta+ \pi j/M$ just before the 
announcement of $j$.

With this modification, every case in the detected events 
are regarded as an outcome of a measurement on 
$\ch_A\otimes\ch_B$. For example, conclusive events 
correspond to the operator
$\hat{R}_{\rm con}\equiv\sum_j (\hat{A}_{0,j}+\hat{A}_{1,j})\otimes
(\hat{B}^{(j)}_0+\hat{B}^{(j)}_1)$, and 
the events with a bit error ($a\neq b$) to 
$\hat{R}_{\rm bit}\equiv \sum_j (\hat{A}_{0,j}\otimes\hat{B}^{(j)}_1
+ \hat{A}_{1,j}\otimes\hat{B}^{(j)}_0)$.
The summation over $j$ implies that these operators are
invariant under the discrete rotations $\Omega_M$.
Hence, these should take the form of 
$\bigoplus_{k=0}^{M-1} \hat{s}_k$, where $\hat{s}_k$ acts on 
the Hilbert space $\ch_k$ spanned by 
$\ket{0}_k\equiv \ket{2k}_A\ket{1}_B$ and $\ket{1}_k\equiv
\ket{2(k+1\;\text{mod}\;M)}_A\ket{-1}_B$,
the states with total angular momentum $2k+1$ (mod $2M$).
Let us introduce the identity $\hat{1}_k$
and Pauli operators $\hat{X}_k\equiv \ket{0}_{k}{}_k\bra{1}+
\ket{1}_{k}{}_k\bra{0}$, $\hat{Z}_k\equiv \ket{0}_{k}{}_k\bra{0}-
\ket{1}_{k}{}_k\bra{1}$ for the qubit $\ch_k$.
Then, we can simply express the operators as 
\begin{eqnarray}
 \hat{R}_{\rm con} &=& M^{-1}
\bigoplus_{k=0}^{M-1} (\hat{1}_k-\cos^2\Theta\hat{X}_k),
\nonumber \\
 \hat{R}_{\rm bit} &=& (2M)^{-1}
\bigoplus_{k=0}^{M-1} (\hat{1}_k-\hat{X}_k).
\label{eq:op-con-err}
\end{eqnarray}

The next step is to describe Bob's measurement $\hat{B}^{(j)}_b$
as a filter followed by 
an ideal measurement on a qubit, as in the security proof of 
the B92 \cite{TKI03}. Consider
a virtual qubit $D$ with $z$ basis $\{\ket{0_z}_D,\ket{1_z}_D\}$
and $x$ basis $\{\ket{0_x}_D,\ket{1_x}_D\}$, where
$\ket{b_x}_D\equiv (\ket{0_z}_D+(-1)^b\ket{1_z}_D)/\sqrt{2}$.
Define an operator $\hat{F}^{(j)}:\ch_B \to \ch_D$ by
\begin{multline}
 \hat{F}^{(j)}=\sqrt{2/M}[\sin(\Theta/2)\ket{1_x}_D 
{}_B\bra{\xi_{\overline{(\Theta+\pi)/2+\pi j/M}}}
\\
+\cos(\Theta/2)\ket{0_x}_D 
{}_B\bra{\xi_{\overline{\Theta/2+\pi j/M}}}
].
\nonumber
\end{multline}
Since 
$\hat{B}^{(j)}_b=\hat{F}^{(j)\dagger}\ket{b_z}_D{}_D\bra{b_z}\hat{F}^{(j)}$
holds, Bob's measurement can be regarded as the filtering process
described by Kraus operator $\hat{F}^{(j)}$, which tells
whether the event is conclusive, followed by $z$-basis measurement 
on qubit $D$, which gives the bit value $b$. 
To prove the security, we are interested in what would happen if
Bob measured the qubit on $x$ basis, and how Alice could predict 
the outcome of that measurement. Bob's $x$-basis measurement 
corresponds to the operator $\hat{B}^{\prime(j)}_b\equiv 
\hat{F}^{(j)\dagger}\ket{b_x}_D{}_D\bra{b_x}\hat{F}^{(j)}$
acting on $\ch_B$. In order to predict the outcome, Alice 
could measure $\{\hat{A}'_{a,j}\}$ instead of $\{\hat{A}_{a,j}\}$,
where 
$\hat{A}'_{a,j}\equiv P(e^{i\varphi}\ket{\xi_{\pi j/M}}_A
-(-1)^ae^{-i\varphi}\ket{\xi_{\Theta+\pi j/M}}_A
)/4$.
Here $\varphi$ is a parameter we can freely choose.
Since $\hat{A}_{0,j}+\hat{A}_{1,j}=\hat{A}'_{0,j}+\hat{A}'_{1,j}$,
this change of measurement does not affect the announcement 
of $j$. The ``phase error'' 
event where Alice's prediction fails ($a\neq b$)
corresponds to the operator 
$\hat{R}_{\rm ph}\equiv \sum_j 
\hat{A}'_{0,j}\otimes\hat{B}^{\prime(j)}_1
+ \hat{A}'_{1,j}\otimes\hat{B}^{\prime(j)}_0$,
which is again rotationally invariant.
After some algebra with
$\varphi'\equiv \varphi + (\Theta/2)$, 
we can express it as 
\begin{multline}
 \hat{R}_{\rm ph}=
\frac{1}{2M}\bigoplus_{k=0}^{M-1} 
\left(
\cos2(k\Theta+\varphi')(\cos^2\Theta\hat{1}_k-\hat{X}_k)
\right.
\\
\left.+(1/2)\sin2\Theta\sin2(k\Theta+\varphi')\hat{Z}_k
\right)
+\hat{R}_{\rm con}/2.
\label{eq:op-ph}
\end{multline}

The rest of the task is to consider how we can estimate 
the number of the phase errors from the actually observed 
quantities. For that purpose, we consider the real protocol 
and a virtual protocol, both of which look identical in 
Eve's point of view. 
Suppose that there are $3N$ detected events for simplicity
(The argument is the same for $N+o(N)$ instead of $3N$).
Alice and Bob randomly group these into three sets of $N$ events.
For the first one, Alice measures the angular momentum 
of system $A$ in the virtual protocol. In the real protocol,
she refrains from disclosing 
$j$ and just discards the events. Let $r_{2k}$ be
the relative frequency (number of events divided by $N$)
of the outcome $2k$.
For the second group, Alice and Bob together measure
the bit error $\hat{R}_{\rm bit}$
in either of the protocols. 
Let $r_{\rm err}$ be its relative frequency.
For the third group, from which they 
try to extract the key, Alice and Bob obtain
$r_{\rm con}N$ conclusive bits in the real protocol.
Let $r_{\rm bit}N$ be the number of errors in these bits.
In the virtual protocol, they measure 
$\hat{R}_{\rm ph}$ and determine its relative frequency
$r_{\rm ph}$.

Thanks to the random grouping, $r_{\rm bit}=r_{\rm err}$
in the limit
of $N\to\infty$ (more precisely, for any $\epsilon_0>0$,
the probability 
of $|r_{\rm bit}-r_{\rm err}|>\epsilon_0$ is exponentially 
small in the limit). Further, 
as in \cite{TKI03}, there must be a state 
$\hat\rho\equiv \bigoplus_k p_k \hat{\rho}_k$, 
where $\hat{\rho}_k$
is a normalized density operator acting on $\ch_k$,
such that $\tr[\hat\rho\ket{2k}_A{}_A\bra{2k}]=r_{2k}$,
$\tr[\hat\rho\hat{R}_{\rm bit}]=r_{\rm bit}$,
$\tr[\hat\rho\hat{R}_{\rm con}]=r_{\rm con}$,
and $\tr[\hat\rho\hat{R}_{\rm ph}]=r_{\rm ph}$.
Let us write $X_k\equiv \tr(\hat\rho_k\hat{X}_k)$
and $X\equiv \sum_k p_k X_k$. The quantity 
$X$ can be determined from the observed 
value $r_{\rm err}$ or $r_{\rm con}$ through the relations
resulting from Eqs.~(\ref{eq:op-con-err}),
\begin{equation}
 2Mr_{\rm bit}=1-X, \;
Mr_{\rm con}=1-X\cos^2\Theta.
\end{equation}
Note that we can omit the measurement of $r_{\rm err}$ unless
$\Theta=\pi/2$.
Using Eq.~(\ref{eq:op-ph}) and 
the relation $X_k^2+\tr(\hat\rho_k\hat{Z}_k)^2\le 1$,
we obtain
\begin{equation}
 r_{\rm ph}\le \frac{r_{\rm con}}{2}+
\frac{1}{2M}\sum_kp_k f_{2(k\Theta+\varphi')}(X_k),
\label{eq:ph-bound}
\end{equation}
where
$$
f_\phi(x)\equiv 
\cos\phi (\cos^2\Theta-x)
+\frac{1}{2}\sin2\Theta|\sin\phi| \sqrt{1-x^2}.
$$
It is seen that if $\cos 2(k\Theta+\varphi')\le 0$,
the phase error contribution from $\ch_k$ is too high
even if $X_k=1$ (no bit errors). We thus should choose
$\varphi'$ such that the contribution from such 
``bad'' subspaces is minimized. Finally, we derive 
an inequality from the fact that Eve cannot touch 
system $A$ directly. Even though Eve may freely choose 
which events should be detected,
$\{r_{2k}\}$ must still satisfy 
$r_{2k}\le \eta_d^{-1}\tr[\hat\rho_{AC} \ket{2k}_A{}_A\bra{2k}]$,
which leads to
\begin{multline}
 p_k(1-\sqrt{1-X_k^2})+ p_{k-1}(1-\sqrt{1-X_{k-1}^2})
\\
 \le 2\eta_d^{-1}\tr[\hat\rho_{AC} \ket{2k}_A{}_A\bra{2k}],
\label{eq:loss-bound}
\end{multline}
where $p_{-1}\equiv p_{M-1}$.
An upper bound $\bar{r}_{\rm ph}$ of $r_{\rm ph}$ 
can thus be calculated from 
the observed quantities $X$ and $\eta_d$ by taking the maximum 
of the rhs (right-hand side) of Eq.~(\ref{eq:ph-bound})
over $\{p_k, X_k\}$, under the constraints of Eq.~(\ref{eq:loss-bound})
and $X=\sum_k p_k X_k$.

Now the situation is summarized as follows.
Alice and Bob have $r_{\rm con}N$ conclusive bits
with $r_{\rm err}N$ errors. Bob's bits can be regarded as
outcomes of $z$-basis measurements on $r_{\rm con}N$ qubits,
and if he had measured those qubits on $x$-basis, Alice 
could have predicted the outcomes with at most $\bar{r}_{\rm ph}N$
errors. Then, by the argument in \cite{Koashi05}, 
we can extract an unconditionally secure secret key of length
$$
G_N=Nr_{\rm con}[1-h(r_{\rm err}/r_{\rm con})-h(\bar{r}_{\rm ph}/r_{\rm con})],
$$
where $h(x)\equiv -x \log_2 x - (1-x)\log_2(1-x)$.
We can also derive the same key rate by assuming virtual 
qubits in Alice's side and invoking Shor-Preskill 
argument \cite{Shor-Preskill00}.

Using the general prescription derived above, here 
we concentrate on the most interesting case, 
the high channel-loss limit ($\eta\to 0$) 
when Alice uses an attenuated
laser source with mean photon number $\mu$,
namely, $\mu_n=e^{-\mu}\mu^n/{n!}$. 
For a fair comparison of various protocols,
we assume that, in addition to the 
transmission $\eta$, the channel 
applies a random rotation of polarization  
with probability $\epsilon$. This results in 
the observed quantities $X=1-\epsilon$ and 
$\eta_d=\mu \eta$ in the limit $\eta\to 0$.
Let $K$ be a positive integer satisfying $K\le M-2$ and 
$2L(K-1)<M$, namely, $\cos(K-1)\Theta>0$.
We will show that by setting $\mu=(\gamma\eta)^{1/K}$,
the key gain of $O(\eta^{(K+1)/K})$ is achievable in 
the high-loss limit $\eta\to 0$, when $\epsilon$ is 
small enough.

Consider the limit $\eta\to 0$ with 
$\mu=(\gamma\eta)^{1/K}$.
For $k=K+1$, 
the rhs of Eq.~(\ref{eq:loss-bound})
converges 
to $\zeta_K\gamma$, where $\zeta_K\equiv 
[2^{K}(K+1)!]^{-1}$ is a constant. 
Hence, when $K=M-2$, we have 
\begin{equation}
   p_{M-1}(1-\sqrt{1-X_{M-1}^2})+ p_{K}(1-\sqrt{1-X_{K}^2})
\le \zeta_K\gamma.
\label{eq:bad-bound}
\end{equation}
When $K\le M-3$, either $p_k=0$ or $X_k=0$ holds
for $k=K+1,\cdots, M-1$ since the 
rhs of Eq.~(\ref{eq:loss-bound}) vanishes 
for $k\ge K+2$. Therefore,
Eq.~(\ref{eq:bad-bound}) still holds, 
and the contributions from $\ch_{K+1},\cdots,\ch_{M-2}$
is not significant. On the other hand, 
there are no bounds for $\ch_0, \cdots, \ch_{K-1}$.
We thus choose $\varphi'=-(K-1)\Theta/2$, 
making these subspaces ``good'' ones.

\begin{figure}[tbp]
\begin{center}
 \includegraphics[scale=0.450]{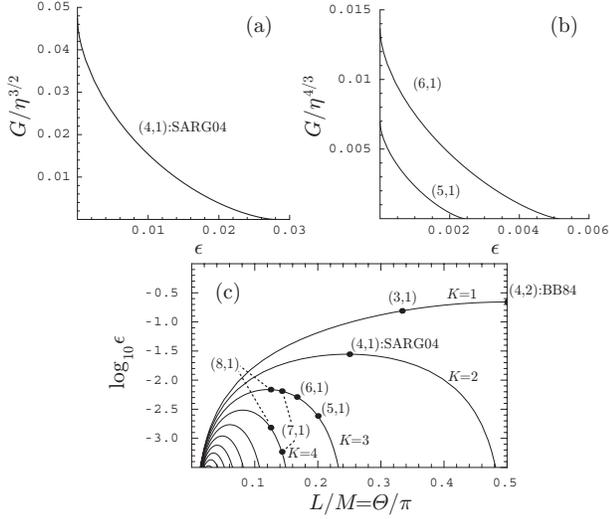}
 \end{center}
 \caption{(a) Key gain $G$ of SARG04 in the high channel loss limit
$\eta\to 0$.
(b) The same for the $(5,1)$ and the $(6,1)$ protocol.
(c) Threshold channel noise $\epsilon$ for achieving
$O(\eta^{(K+1)/K})$ scaling of the key gain.
\label{figure}}
\end{figure}

Let us divide $X$ into two contributions as
$X=qX'+(1-q)X''$, where the first term is
from $\ch_K$ and $\ch_{M-1}$, namely, 
$q\equiv p_K+p_{M-1}$ and $qX'\equiv p_{K}X_K+p_{M-1}X_{M-1}$.
Then, from Eq.~(\ref{eq:bad-bound}), we have 
\begin{equation}
   q(1-\sqrt{1-X^{\prime 2}})
\le \zeta_K\gamma,
\label{eq:q-bound}
\end{equation}
and Eq.~(\ref{eq:ph-bound}) becomes
\begin{equation}
M(2r_{\rm ph}-r_{\rm con})\le q f_{(K+1)\Theta}(X') +(1-q) f(X''), 
\label{eq:ph-bound-limit}
\end{equation}
where 
$f(x)$ is defined as the boundary of 
the convex hull of the union of regions 
below $f_{(K-1)\Theta}(x)$, $f_{(K-3)\Theta}(x)$, $\ldots$,
$(f_\Theta(x)$ or $f_0(x))$. 
For the relevant case with small bit errors ($x$ being close to 1), 
$f(x)$ coincides with $f_{(K-1)\Theta}(x)$.
Let $g(\gamma,\epsilon)$ be the maximum of 
the rhs of Eq.~(\ref{eq:ph-bound-limit}) over 
$q, X', X''$ under Eq.~(\ref{eq:q-bound}) and 
$1-\epsilon=qX'+(1-q)X''$. 
The key gain $G$
(per pulse) in the limit $\eta\to 0$ is given by
\begin{multline}
 G/\eta^{(K+1)/K} \sim \gamma^{1/K}\beta(\epsilon)M^{-1}
\\
\times\left[1-h\left(\frac{\epsilon}{2\beta(\epsilon)}
\right)
-h\left(\frac{1}{2}
+\max\{0, \frac{g(\gamma,\epsilon)}{2\beta(\epsilon)}\}\right)\right],
\label{eq:keyrate}
\end{multline}
where $\beta(\epsilon)\equiv M r_{\rm con}= 1-\cos^2\Theta(1-\epsilon)$.
Since $g(0,0)=f_{(K-1)\Theta}(1)<0$, the rhs of Eq.~(\ref{eq:keyrate}) 
is positive for sufficiently small $\gamma$ and $\epsilon$. 
For a fixed $M\ge 4$, $K$ can be as large as $\lceil M/2\rceil$
when $L=1$. Hence, the key gain scales as $O(\eta^{3/2})$ in 
the SARG04 protocol. The gain after optimization over $\gamma$
is shown in Fig.~\ref{figure}(a). When $\epsilon=0$, the 
optimum intensity of Alice's source scales as $\mu\sim 1.51 \eta^{1/2}$.
While the $(5,1)$ protocol 
and the $(6,1)$ protocol both achieve $O(\eta^{5/4})$ scaling, 
the latter protocol is better as shown in Fig.~\ref{figure}(b).
In addition, we can double the key gain of  
the protocols with even $M$ by separately collecting 
and processing the events where $D_2$ registers no photons 
and $D_1$ registers one photon. 
Fig.~\ref{figure}(c) shows the threshold noise $\epsilon$
below which the key gain of $O(\eta^{(K+1)/K})$ is 
achievable. It is seen that the requirement for the noises
becomes tighter as the exponent improves. 
This threshold depends only on $\Theta$ and not 
on $M$, since the only dependence on $M$ of the rate (\ref{eq:keyrate})
is the $M^{-1}$ factor. The threshold is highest for approximately
$\Theta\sim \pi/4(K-1)$, which is achieved by the 
$(4(K-1),1)$ protocol.

In summary, we have proved 
the unconditional security of the SARG04 protocol 
and shown that the key gain scales as $O(\eta^{3/2})$.
A natural generalization was given for the $(M,L)$
protocols with $M$ linearly polarized states.
When the channel noise is low, the $(M,1)$ protocol
can achieve the key gain of 
$O(\eta^{(K+1)/K})$ with $K=\lceil M/2\rceil$.
One might wonder why we do not achieve 
$K=M-2$, which is the bound due to the USD
(unambiguous state discrimination) attack \cite{Tamaki-Lo04,Chefles98}.
But $K=\lceil M/2\rceil$ is indeed optimal, 
because there is an attack which is a
kind of mixture of USD and PNS. For example, 
if Alice emits 4 photons in the $(6,1)$ protocol,
Eve may apply a filter to the excess 3 photons 
to obtain state $[e^{3i\theta}(\hat{a}^\dagger_{-1})^3
+e^{-3i\theta}(\hat{a}^\dagger_{1})^3]\ket{vac}$ 
with a nonzero probability. She sends the 
remaining one photon to Bob only when the filtering
has been successful. It is then obvious that Eve can 
always determine the bit $a$ after the announcement of $j$.
It is not difficult to extend this attack to 
larger values of $M$. One possible way to avoid 
this kind of attack is to mix the protocols 
with different values of $L$, which is just a
modification of the classical communication part.

The author thanks N.~Imoto and K.~Tamaki for helpful 
discussions. This work was supported by a MEXT Grant-in-Aid 
for Young Scientists (B) 17740265.

\bibliographystyle{apsrev}

\end{document}